\documentclass{article}
\pdfoutput=1
\usepackage{graphicx}
\usepackage{epsfig}
\usepackage{amssymb}
\usepackage{amsmath}
\usepackage{upgreek,bm}
\usepackage{authblk}
\usepackage{float}
\usepackage{color}

\begin{document}

\title{Big Bang Bifurcations in the Tantalus Oscillator under Biphasics Perturbations.} 

\author{H. Arce} 
\author{A. Torres} 
\author{A. Cabrera}
\author{M. Alarc\'on}
\author{C. M\'alaga}
\affil{Departamento de F\'isica, Facultad de Ciencias, Universidad Nacional Aut\'onoma de M\'exico, Apartado Postal 70-542, 04510 M\'exico, Distrito Federal, M\'exico.
}

\maketitle


\renewcommand{\thefootnote}{\arabic{footnote}}

\begin{abstract}
The Tantalus Oscillator is a non linear hydrodynamic oscillator with an attractive limit cycle. In this study we pursue the construction of a biparametric  bifurcation diagram for the Tantalus Oscillator under biphasics perturbations. That is the first time that this kind of diagram is built for this kind of oscillator under biphasics perturbations. Results show that biphasic perturbations have no effect when the coupling time is chosen over a wide range of values. This modifies the bifurcation diagram obtain under monophasics perturbations. Now we have the appearance of periodic increment Big Bang Bifurcations. The  theoretical results are in excellent agreement with experimental observations.

\end{abstract}

\section{Introduction}

Non linear oscillators with an attractive limit cycle, that is to say they exhibit stable periodic oscillations, appear in a wide variety of natural phenomena and artificial devices. The understanding of their behavior when perturbed, how they return to their periodic oscillations, is a matter of great practical importance. When perturbed by a train of brief pulses their response can, under certain conditions, be studied using Resetting Theory \cite{Glass1988,Glass2003}. Resetting Theory sparkled interest in relation to biological systems and medical applications related to the heart beat, breathing, circadian cycles, neuronal oscillations, etc. The theory requires the system to be stable during many oscillation periods and to quickly return to its limit cycle when perturbed, making it unsuitable for many biological systems that seldom behave in this manner \cite{Krogh}. 

Nevertheless, the Tantalus Oscillator has proven to be a good experimental model for Resetting Theory \cite{Chialvo,Arce}. Recent studies have shown that the Tantalus Oscillator exhibits border collision bifurcations and possibly period increment Big Bang Bifurcations (BBB) under monophasic perturbations, a perturbation where a system variable is suddenly increased or reduced \cite{Arce}. In the present study, we explore the response of the Tantalus Oscillator under biphasic perturbations, where a sudden increase of a system variable is immediately restored to its starting value. 

Although oscillators with a stable limit cycle subject to monophasic perturbations have been widely studied \cite{Santillan,Guevara,Galan,Keener}, little is known of their response to biphasic perturbations and there is no information of their bifurcation diagram \cite{Hortensia,Croisier}, despite the fact that biphasic perturbations are present in important practical applications such as implantable cardioverter defibrillators \cite{Kanj}. Therefore, our interest in the Tantalus Oscillator is twofold: first, the construction of the biparametric bifurcation diagram under biphasic perturbations; and second, the comparison with the response to monophasic perturbations.

In the present study we found that the Tantalus Oscillator is insensitive to biphasic perturbations over a wide range of values of the perturbation time ($T_c$), which corresponds to a Phase Transition Curve (PTC) given by the identity function over these intervals. Nevertheless, 
The global bifurcation diagram is completely modified with respect to the monophasic case and period increment BBBs appear. The appearance of BBBs is particularly interesting as they rearrange the bifurcation diagram in their vicinity. Preliminary observation on a saline oscillator suggest that some of the observed phenomena on the Tantalus oscillator are a common feature.

In the following section we present the details of experimental setup and data acquisition, as well as the comparison with the behavior predicted by the hydrodynamic equations. Results are presented in section 3, where PTC and diagrams are constructed. Section 4 has the discussion of the results and the bifurcation phenomena. We summarize our work in the last section.

\section{Experimental setup}

\begin{figure}[t]
\centering
\includegraphics[width=0.9\columnwidth]{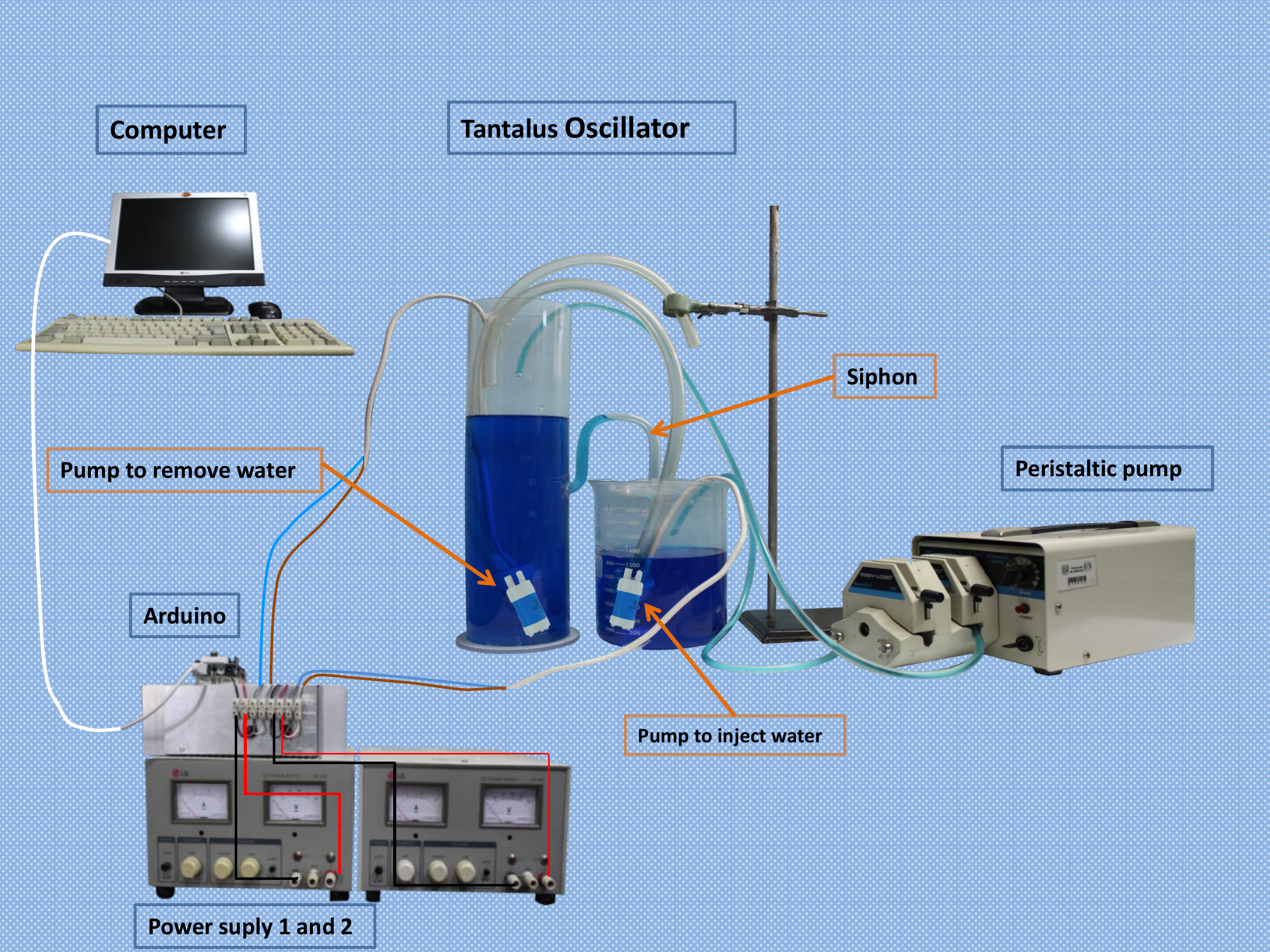}
\caption{Experimental setup. At the center is the Tantalus device, the cylindrical vase with a siphon at mid height and partially filled with blue water. A peristaltic pump injects water at a constant rate to the Tantalus. Two pumps, one submerged in the Tantalus and the other in the reservoir beside it at the outlet of the siphon, are used to inject and subtract water to produce the biphasic perturbations. Pumps are driven by an arduino controlled by a computer.}
\label{fig1}
\end{figure}

A sketch of the experimental setup is presented in figure \ref{fig1}. The main component is the Tantalus device, a $39 cm$ tall cylindrical vase of $9.5 cm$ of diameter, with a siphon  of $0.9 mm$ of diameter located at $17 cm$ from the bottom. A Masterflex L/S peristaltic pump injects tinted water to the Tantalus vase at a constant rate of $9.8$ milliliters per second.
The water level oscillation is produced by the influx provided by the Masterflex pump and the discharge produced by the siphon every time the level reaches the siphon's maximum height. This oscillation displaces a volume $V_0$ of $580$ milliliters of water.

Two additional submersible Comet Elegant pumps of $24$ volts are use to produce the biphasic perturbations. One Comet pump rests at the bottom of an additional water reservoir that collects water from the siphon outlet, it injects a chosen volume of water (a small fraction $I$ of $V_0$) to the Tantalus during a period of $2$ seconds. The second Comet pump, at the bottom of the Tantalus vase and well below the siphon inlet, extracts the same volume of water during the same amount of time, producing a symmetrical biphasic water level perturbation.  Two power supply units LG-GP503 and an arduino controlled with a computer are used to drive the submersible pumps and produce perturbations in the form of isolated pulses or of a train of pulses of a given periodicity. 

\begin{figure}[t]
\centering
\includegraphics[width=0.9\columnwidth]{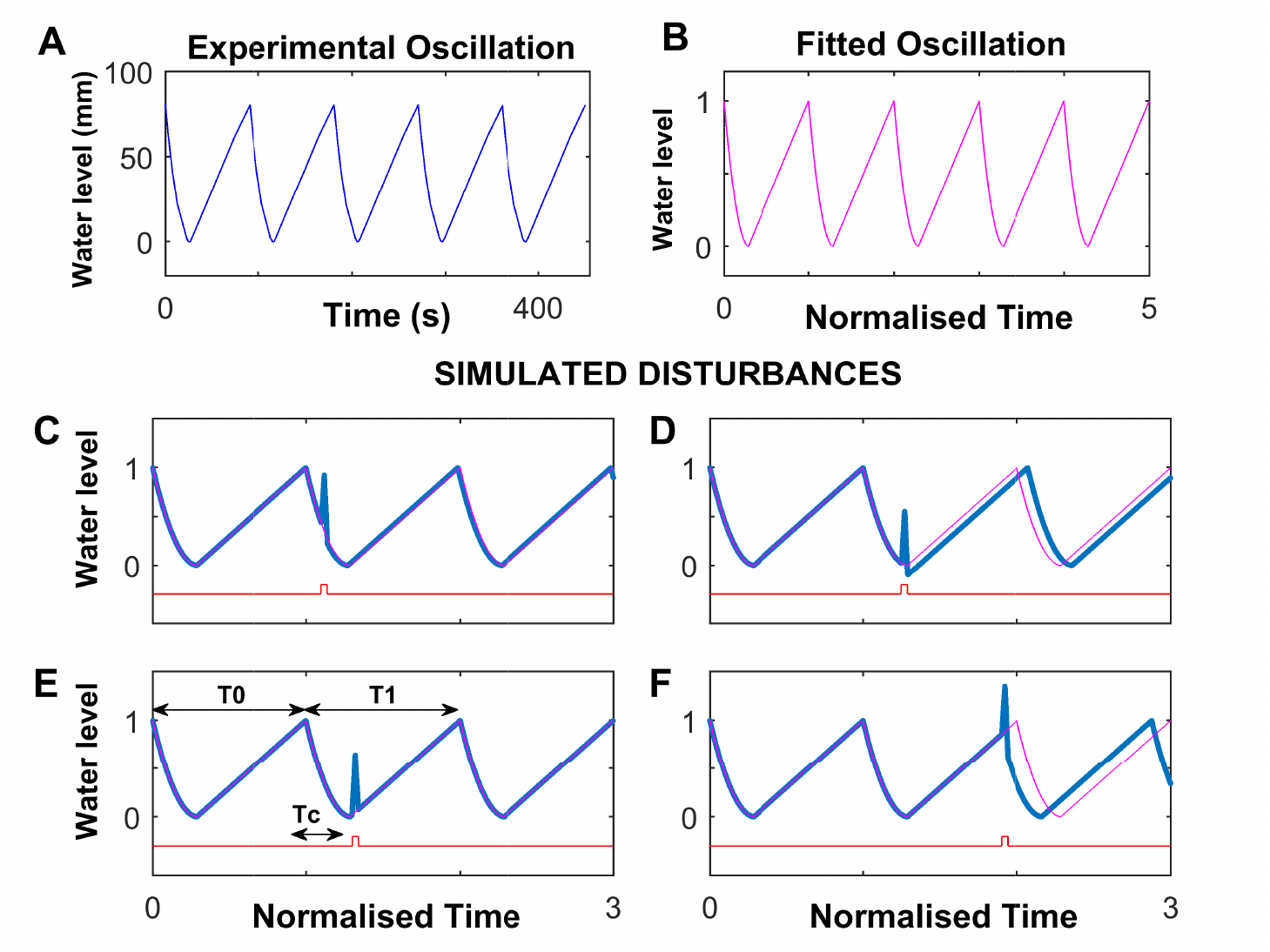}
\caption{Experimental measurements and perturbation simulations. In (A) the measured unperturbed water level oscillation is shown. The oscillation predicted by the model formula \ref{bajada} and  \ref{subida}  over the time normalized with $T_0$ appears in (B). Sub-figures (C) to (F) show the simulated effect of a single perturbation at different times (marked by the red pulse), on the water level behavior (in blue). The pink curve in (C), (D), (E) and (F) represent the unperturbed oscillation.}
\label{fig2}
\end{figure}



\section{Results}

\subsection{Unperturbed oscillations and perturbation simulations}

As done in a previous work \cite{Arce}, we first filmed the unperturbed Tantalus oscillation.
The time evolution of the water was obtained from the video images using Imagej (an open source image processing program). The data of $5$ oscillations is shown in figure \ref{fig2}(A). 
Water level height was measured from the siphon inlet.
We observed an average period of $89 \pm 0.5 s$, where the siphon discharge took $27 s$ and the water refill of the Tantalus lasted $62 s$.

Scaling time with the oscillation period and height with the maximum water level, a model for the water level as a function of time $h(t)$ can be constructed as the repetition of a quadratic drop, obtained from the Bernoulli hydrodynamic equation for the siphon, followed by the linear grow imposed by the peristaltic pump \cite{brazuca}. The equation describing the water level descend is

\begin{equation}
N(t) = 11.47 \ t^2 - 6.76 t + 0.99,
\label{bajada}
\end{equation}

\noindent while the water level increase is given by

\begin{equation}
N(t) = 1.4 \ t - 0.41.
\label{subida}
\end{equation}

The model equation can be manipulated to simulate the perturbations and predict the Tantalus response. 
Considering the beginning of the oscillation when the water is at its highest, the perturbation at time $T_c$ is simulated by increasing the value of $h$ in small steps, and decreasing in a symmetrical way. 
Between steps the system evolves in time for a fraction $10^{-4}$ of the duration of the perturbation
The simulated pulse (the perturbation) lasts about $\% 4$ of the oscillation period.

\begin{figure}[h]
\includegraphics[width=0.9\columnwidth]{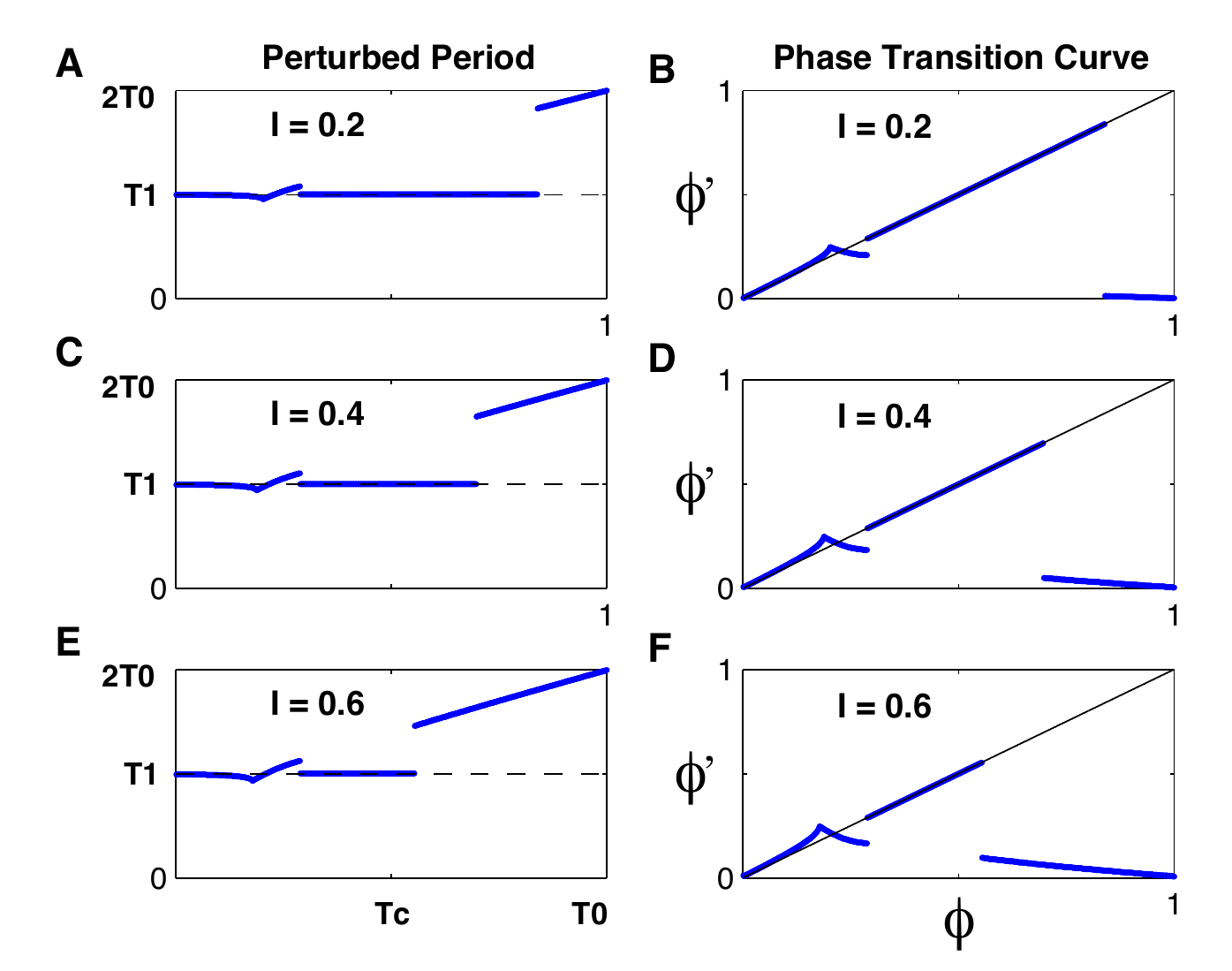}
\caption{Perturbed periods and PTC. In (A), (C) and (E) the perturbed period $T_1$ vs $T_c$ is shown for perturbation intensities $I=0.2$, $0.4$ and $0.6$, respectively. A thousand values of $T_c$, uniformly distributed in the interval $[0,T_0]$, where explored. PTC are presented in (B), (D) and (F).}
\label{fig3}
\end{figure}

Figures \ref{fig2}(C) and (D) show, in blue, the effect of perturbing close to the beginning and the end of the water discharge produced by the siphon.
The unperturbed oscillation is shown in pink for comparison, and the square pulse is shown in red.
The perturbation applied at the beginning of the siphon discharge, figure \ref{fig2}(C), has a negligible effect on the oscillation. 
On the contrary, a perturbation close to the end of the discharge delays the oscillation, see figure \ref{fig2}(D). 
Similarly, a perturbation at the beginning of the Tantalus refill has no effect, while at the end advance the oscillation, see figures \ref{fig2}(E) and (F). 
In figure \ref{fig2}(E), the time intervals used in this work are indicated. $T_0$ corresponds to the unperturbed period, and $T_1$ is the oscillation period modified by the perturbation.


\subsection{Phase Transition Curves}

Let $T_0$ be the unperturbed oscillation period and $T_1$ the period modified by the biphasic perturbation at time $T_c$. In figure \ref{fig3}(A), the computed values of $T_1$ vs $T_c$ for a thousand simulations, of a single perturbation of intensity $I = 0.2$, are presented in blue. The black dashed line represent the reference value $T_1 = T_0$. The two intervals where the perturbation has no effect (where $T_1 = T_0$) correspond to values of $T_c$ beyond the maximum and minimum values of the water level oscillation. Figures \ref{fig3}(C) and (E) show results of simulations with intensities $I=0.4$ and $0.6$, respectively. Naturally, the intervals where $T_1 = T_0$ shrink as intensities grow.

Figures \ref{fig3}(B), (D) and (F) are the PTC related to the simulations in figures \ref{fig3}(A), (C) and (E), respectively. PTC are needed to construct the bifurcation diagram and compare the phase when the system was perturbed $\phi= T_c/T_0$ with the phase shifted after the perturbation 
$\phi'= [T_0 -(T_1-T_c)]/T_0 \ (\mathrm{mod}\ 1)$ \cite{Glass1988,Glass2003,Arce}. 
As observed with monophasic perturbations, there are two apparent discontinuities in the PTC. The first appears at $\phi \simeq 0.27$ in all cases (figures \ref{fig3}(B), (D) and (F)), and correspond to the end of the siphon discharge and the beginning of the Tantalus refill. The position of the second discontinuity depends on the perturbation intensity. Notice that, in contrast with the monophasic PTC made of straight lines, biphasic PTC can have slightly curved sections.

\begin{figure}[t]
\centering
\includegraphics[width=0.9\columnwidth]{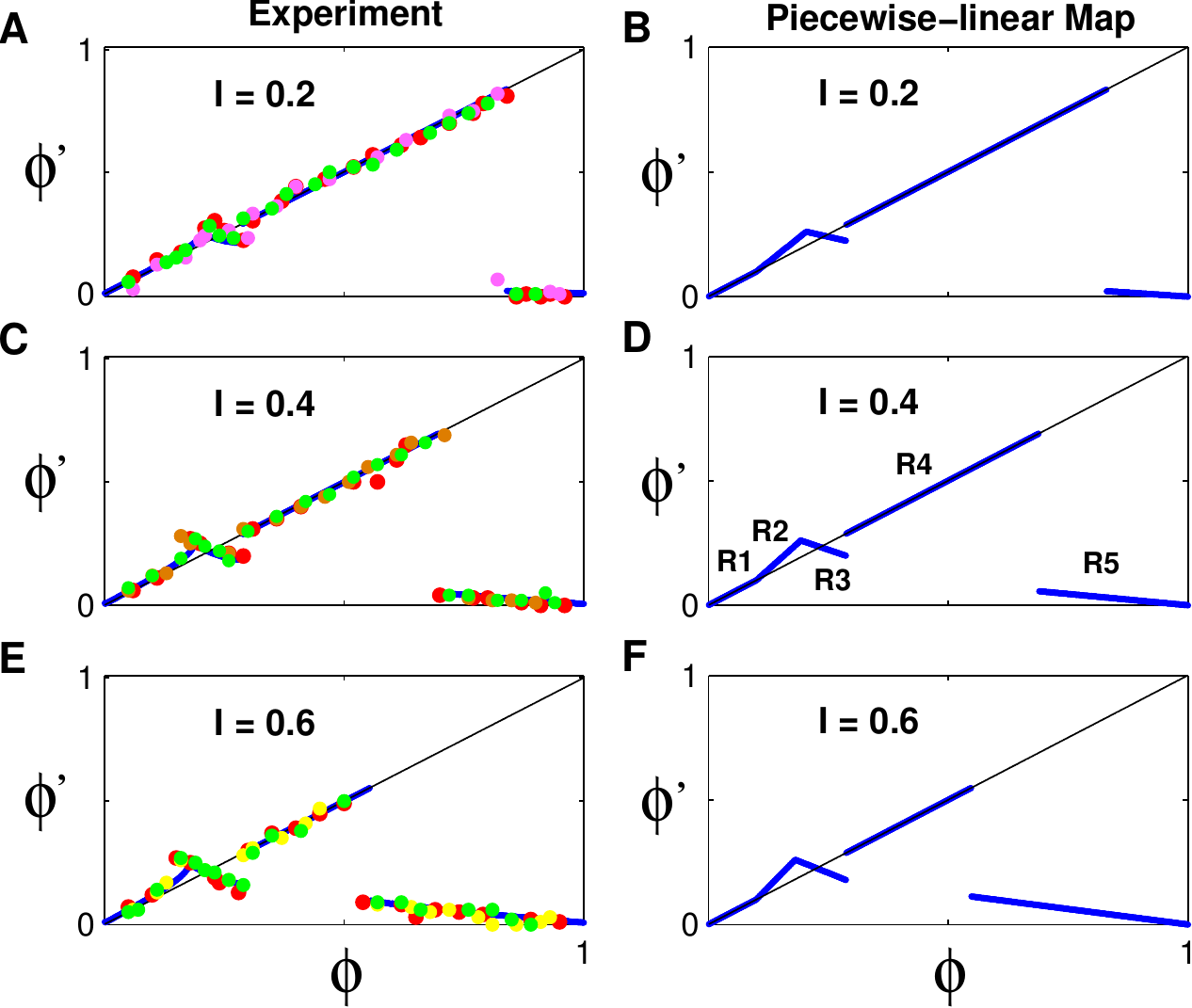}
\caption{PTC experimental results (left) and piecewise fit (right). Experimental data in different colors represent different experimental realizations and are plotted on top of the theoretical fit (left). Plots as the top, middle and bottom correspond to $I=0.2$, $0.4$ and $0.6$ respectively.}
\label{fig4}
\end{figure}

Experiments confirm the obtained PTC, as plotted on the left of figure \ref{fig4}. Three sets of experiments (shown in different colors) for each intensity value are plotted on 
top of the theoretical curves of figure \ref{fig3}. Results show remarkable agreement with experiments. Experimental data suggest the existence of real discontinuities.



The curved sections of the PTC, mentioned above, result on a very time consuming computation of the Biparametric Bifurcation Diagram, due to the need of interpolations for the description of these curved sections. That is why we introduced an approximate piecewise linear PTC, show in figure \ref{fig4}(B), (D) and (F), in which the straight segments are named, from the left, $R1$ to $R5$. The use of piecewise linear PTC reduces the time for computing the Biparametric Bifurcation Diagrams in three orders of magnitude.

\subsection{Analytic results and bistability}

Beyond the effect of isolated perturbations or pulses, the PTC are used mainly to predict the system behavior under periodic perturbations. Let $\tau$ be the period of a train of pulses that represent the perturbation, and $\phi_0$ the phase when the first of these pulses perturbs the system. If $g(\phi )$ is the effect or shift produced by a pulse at $\phi$, the sequence of shifted phases produced by the train of pulses can be written as

\begin{equation}
\phi_{i+1} = g( \phi_i) + \tau \ (\mathrm{mod}\ 1).
\end{equation}

\begin{figure}[t]
\centering
\includegraphics[width=0.9\columnwidth]{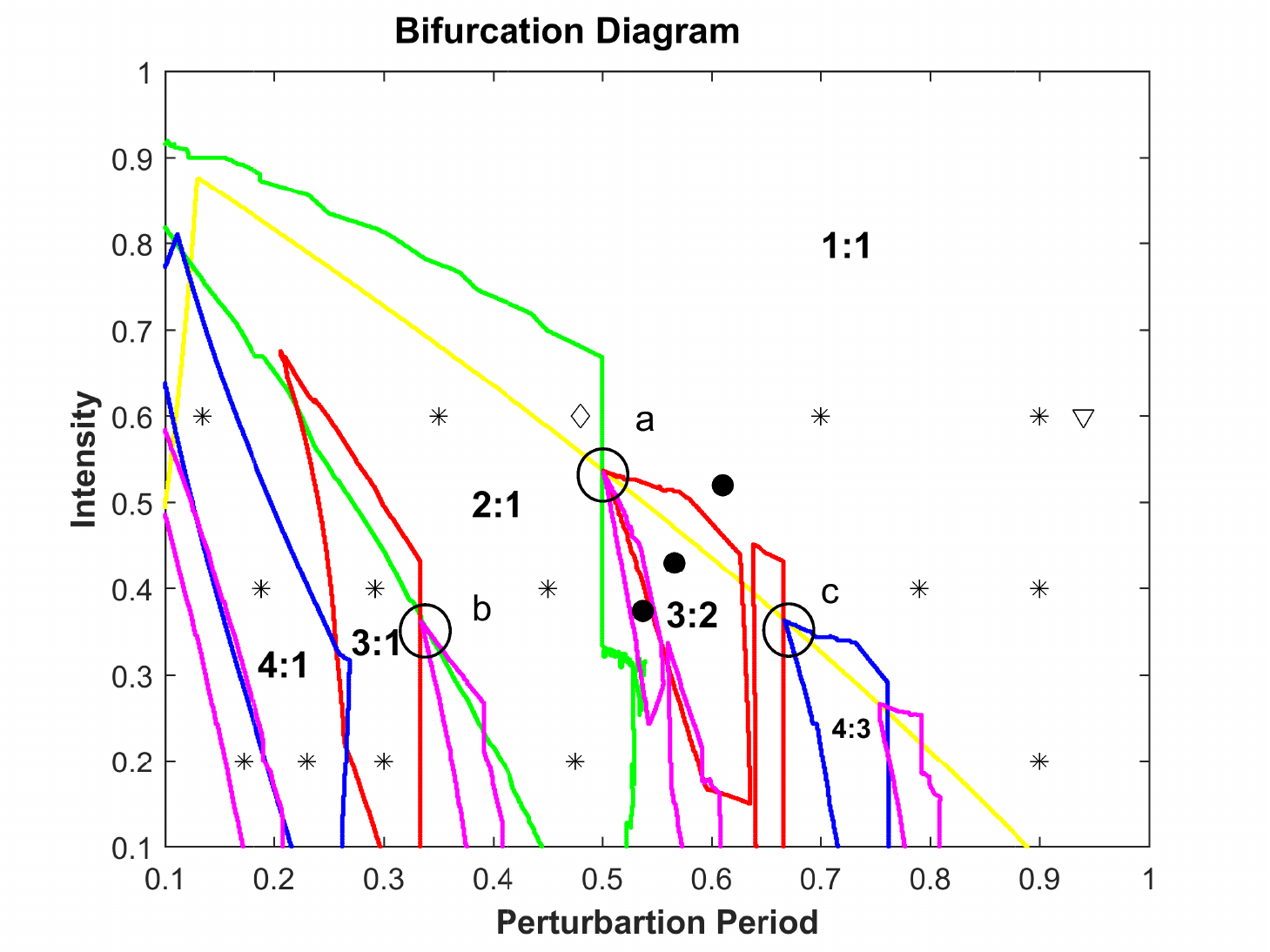}
\caption{Bifurcation Diagram. Colored lines enclose regions where a given rhythm was observed. The rhythm $1:1$ was observed over the yellow line. Intersection between regions correspond to observed bistabilities. Some of the observed rhythms are indicated in the diagram: asterisks indicate experimentally verified rhythms, the triangle corresponds to the auto bistability of figure \ref{fig5}, the diamond corresponds to the auto bistability of figure \ref{fig6}, open circles show BBBs, and closed circles correspond to an experimentally verified increment increment bifurcation sequence of the BBB marked with an $a$.}
\label{fig8}
\end{figure}

Using this relation a thousand times, we found the different behaviors shown in the Global Bifurcation Diagram in figure \ref{fig8}. Known as coupling rhythms or simply rhythms $M:N$, the system behavior consists of its reaction to a train of pulses where $M$ applied pulses fit in $N$ perturbed oscillations.

Some rhythms can be found directly from the analytic expressions, without this procedure, noticing that if the PTC intersects the identity with a slope smaller than one in absolute value then we have a stable attractor \cite{devaney}. In this way, all stable attractors can be found analytically as we know the dependence of the PTC segments on the intensity and period of the perturbations. In these points, the system is perturbed always at the same phase for each period $\tau$, corresponding to a rhythm $1:1$. All points above the yellow line in figure \ref{fig8} correspond to the rhythm $1:1$. 

\begin{figure}[t]
\centering
\includegraphics[width=0.9\columnwidth]{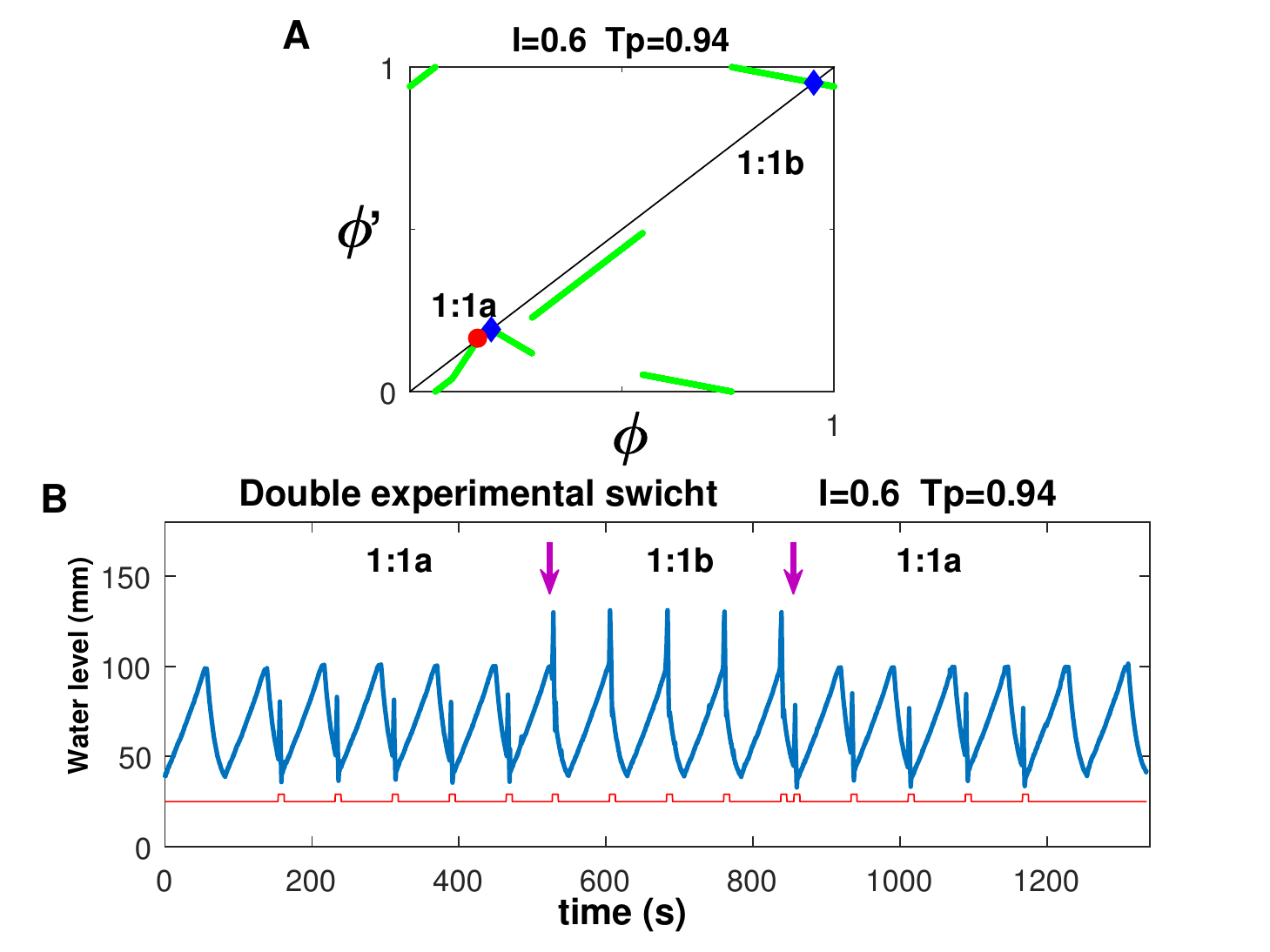}
\caption{Theoretical bistability and double switch between rhythms $1:1a$ and $1:1b$. Figure A shows that perturbations with intensities of $0.6$ and $T_p = 0.94$ have two attractors of rhythm $1:1$ (named $1:1a$ and $1:1b$). In figure B, arrows indicate the times where a small variation on the perturbation times where applied to produce the switch between rhythms. These observations correspond to the triangle in the bifurcation diagram of figure \ref{fig8}.}
\label{fig5}
\end{figure}

Another result can be found analytically, that to our knowledge has been reported only in two previous publications related to biphasic perturbations \cite{Hortensia,Croisier}. Figure \ref{fig5}(A) shows a PTC of intensity $0.6$ and $\tau = 0.94$ that crosses the identity at three points. 
Two of these points correspond to stable attractors (the blue points in figure \ref{fig5}(A)), while the third is unstable (the red point in figure \ref{fig5}(A)). We have called $1:1a$ the rhythm in the segment $R3$ and $1:1b$ the rhythm in the segment $R5$. The experimental observation of the bi-stability is shown in figure \ref{fig5}(B). 
Figure \ref{fig5}(B) shows, for the first time, a double switch between two bistable rhythms within a single experiment, produced by two slight changes in the coupling intervals. The first pulse is applied at $\phi=0.19$, corresponding to the fixed point in segment $R3$ of figure \ref{fig5}(A). During five pulses of period $0.94$ times the natural period of oscillation, the rhythm $1:1a$ is observed. 
The sixth pulse is applied at a different period (indicated by a magenta arrow) to correspond with the phase $0.95$, the fixed point in the segment $R5$ of figure \ref{fig5}(A). 
During the next five pulses of period $0.94T_0$ the rhythm $1:1b$ was observed. The second magenta arrow indicates a new change in the coupling period to recover the perturbations at phase $0.19$ and the $1:1a$ rhythm. 
Notice that the switch between the desired rhythms required a single experimental manipulation and no transient times, proving the Tantalus oscillator to be an adequate experimental model for reseting theory.

\begin{figure}[t]
\centering
\includegraphics[width=0.9\columnwidth]{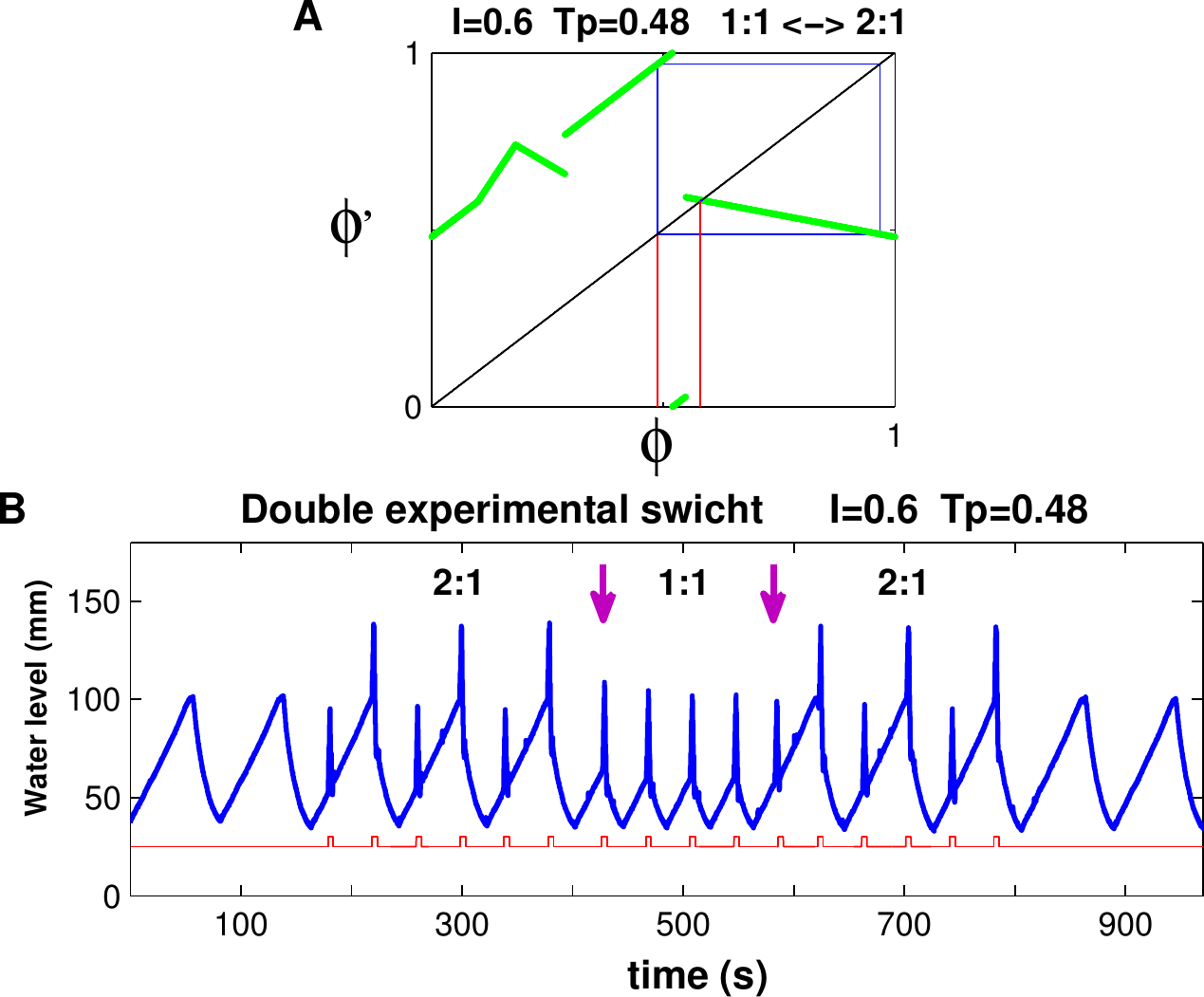}
\caption{Theoretical bistability and double switch between rhythms $1:1$ and $2:1$. Figure A shows that perturbations with intensities of $0.6$ and $T_p = 0.48$ have two attractors of rhythms $1:1$ and $2:1$. In figure B, arrows indicate the times where a small variation on the perturbation times where applied to produce the switch between rhythms. These observations correspond to the diamond in the bifurcation diagram of figure \ref{fig8}.}
\label{fig6}
\end{figure}

Another bistability, not present in the monophasic case, can be found analytically. Figure \ref{fig6}(A) shows two orbits produced by perturbations of intensity $0.6$ and period $0.48$ but different initial conditions. 
One is stable with a single fixed point and corresponds to a rhythm $1:1$, the other, enclosing the first one, corresponds to a rhythm $2:1$. 
This type of bistability is common, with an external orbit of arbitrary periodicity.
Figure \ref{fig6}(B) shows how these two rhythms were obtained in a single experiment. 
Starting with a rhythm $2:1$, the first magenta arrow shows the time when the coupling period was changed to obtain the rhythm $1:1$ during four perturbations before recovering the rhythm $2:1$.

\subsection{Experimental Bifurcation Diagram and BBB}

Although it is possible to find analytically orbits of periodicity higher than those studied in the previous section, it is easier to compute them numerically. 
We have performed these computations using intensities from $0.1$ to $1$ in steps of $0.001$, and periods from $0.1$ to $1$ also in steps of $0.001$. 
In each case, ten different initial phases were considered, from $0$ to $0.9$ in steps of $0.1$. 
Periodicities up to $16$ where found after performing $2560$ iterations in each case.

\begin{figure}[t]
\centering
\includegraphics[width=0.9\columnwidth]{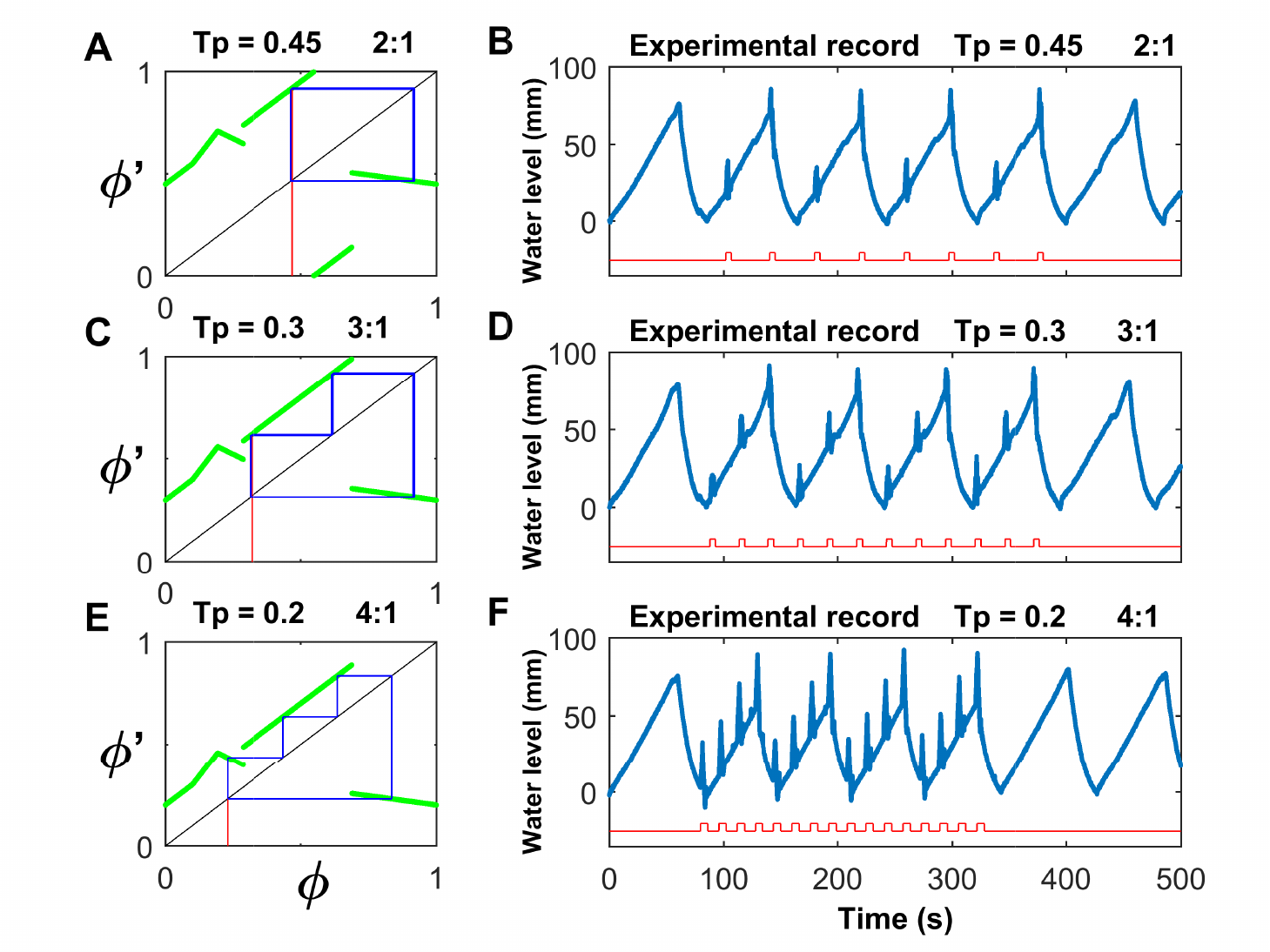}
\caption{Coupling of periodic perturbations (left) and the Tantalus perturbed oscillations for intensities of $0.4$ (right). Graphs on the left show how iterations over the PTCs predict the rhythms $2:1$, $3:1$ and $4:1$. On the right, the corresponding perturbation experiments with biphasic pulses of $232 \ ml$ show a good agreement. Top, middle and bottom rows correspond to $T_p = 0.45,$ $0.3,$ and $0.2$ respectively.}
\label{fig7}
\end{figure}

Three typical orbits, related to intensities $0.4$, are shown in figure \ref{fig7}. 
Figure \ref{fig7}(A) corresponds to a perturbation period of $0.45$ starting at phase $0.47$ indicated by the vertical red line. Notice the orbit hits twice the PTC at two different phases, which means that with every pulse the oscillator will be placed consecutively in each of the phases. 
In figure \ref{fig7}(B) can be found the corresponding experimental observation, notice the rhythm is reached immediately, without a transient behavior. 
Figures \ref{fig7}(C) and (E) show the effects of perturbations of the same intensities, but periods $0.3$ and $0.2$ respectively, and different starting phases indicated by the red line (experiments are shown in figures \ref{fig7}(D) and (F)). Stable orbits were found with periodicities of $3$ and $4$ respectively, produced by a ``channel" in the PTC formed by the segments $R1$ to $R4$ and the identity below them (in black), where the orbit is trapped for some time. 
The width of the ``channel" shrinks as the perturbation period is shortened, making the orbit bounce more times inside it before leaving, only to be redirected back to it by the segment $R5$. 
This feature can be found in the case of monophasic perturbations \cite{Arce,avrutin,perez}.

The bifurcation diagram in figure \ref{fig8} shows rhythms with periodicity up to $5$. 
Although we found higher periodicity rhythms, these are not shown in the Bifurcation Diagram for the sake of clarity. 
Regions bounded by curves of the same color share the same periodicity. 
Four regions of periodicity $5$ can be found in magenta, corresponding to rhythms $5:1$, $5:2$, $5:3$ and $5:4$, from left to right. Rhythms $1:1$ where found above the yellow curve.

\begin{figure}[H]
\centering
\includegraphics[width=0.9\columnwidth]{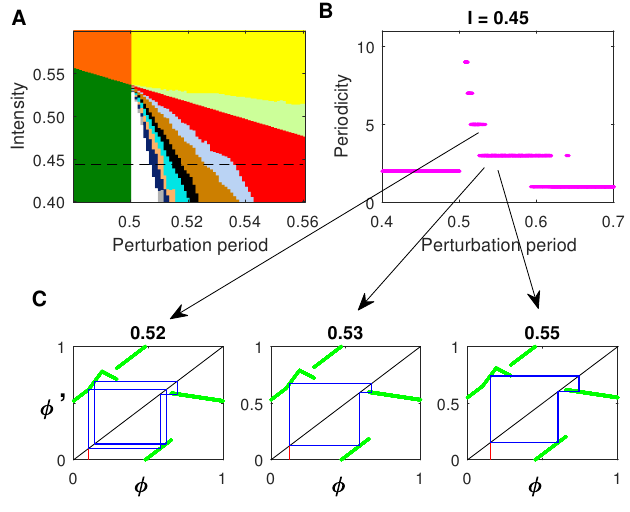}
\caption{Period increment BBB. Subfigure A shows a close up view of the BBB point marked with an $a$ in the bifurcation diagram of figure \ref{fig8}. Colored regions define domains of the bifurcation diagram where a specific periodicity is observed. In the yellow and green regions rhythms of periodicities $1$ and $2$ are observed, while orange corresponds to a bistable region between these rhythms. Red corresponds to $3$ and light green in between red and yellow the bistable region of $1$ and $3$ periodicities. Brown corresponds to $5$ and gray to the bistable $3$ and $5$, and so on. Periodicities at the dashed line at intensity $0.45$ are plotted in B. Notice that periodicities increase by two when approaching the value of $0.5$ from above, and overlap in bistable regions. The sequence in C shows how border collision produces the bifurcation between periodicities of $3$ and $5$. As the perturbation period goes from $0.55$ to $0.52$ the orbit crosses the discontinuity between segments $R4$ and $R5$ (see figure \ref{fig4}D).}
\label{fig9}
\end{figure}

Notice that high periodicity regions are placed to the left, present many bistable intersections, and do not show accumulation point, in contrast with the monophasic case. 
Experiments are shown with asterisks, where produced with intensities $0.2$, $0.4$ and $0.6$, and in all cases confirmed the predicted rhythms. The bistabilities in figures \ref{fig6} and \ref{fig5} correspond to the points represented by a diamond and a triangle, respectively.
The most interesting feature found in the Bifurcation Diagram is the possible existence of period increment Big Bang Bifurcations (BBB). 
The circles in figure \ref{fig8} indicate three points in the Bifurcation Diagram where BBB may exist. 
In the next section we discuss why one of these points must be a BBB. 
The point tagged with the letter $a$ corresponds to rhythms $1:1$, $2:1$, $3:2$ and $5:3$. 
A closer look around this point can be found in figure \ref{fig9}(A) where higher periodicity regions are included.

The regions of rhythms $1:1$ and $2:1$ are in yellow and dark green respectively, and their bistable region is painted in orange. 
The vicinity of the BBB point is a sort of pie chart with an increasing number of slices at a given angle (at the white slice). 
The red slice corresponds to the rhythm $3:2$, or periodicity $3$. 
A light green slice of bistable periodicities $1$ and $3$ is found above the red slice, with a gray bistable slice of periodicities $3$ and $5$ below it. 
A region of periodicity $5$ is colored in brown, with an adjacent black slice of bistable periodicity $5$ and $7$. 
Slices of increasing periodicity are arranged in the same manner, increasing in steps of two in between bistable slices.

Figure \ref{fig9}(B) shows part of the one-parameter bifurcation diagram for perturbations of intensity $0.45$, shown with a dashed red line in figure \ref{fig9}, obtained varying the perturbation period form $0.4$ to $0.7$ in the horizontal axis. The vertical axis show the periodicities. From the right, periodicities start from $1$ and increase in steps of two with overlaps corresponding to bistable regions. As periodicities increase, the steps become shorter. Figure \ref{fig9}(C) shoes how this periodicity change happens. 
Starting from the left, three orbits of rhythms with periodicities $5$, $3$ and $3$ are shown, obtained with perturbation periods of $0.52$, $0.53$ and $0.55$ respectively. The orbit on the right bounces three times off the PTC, as does the one in the middle but closer to the discontinuity between segments $R5$ and $R4$. As the perturbation period decreases, the orbit ends up avoiding the segment $R5$. Then, a bifurcation occurs and an orbit of periodicity $5$ appears, with an extra turn around the PTC and two more bounces. As with monophasic perturbations, this is a border collision bifurcation, as it happens when the orbit approaches a discontinuity. If we further decrease the period, a new bifurcation appears when the orbit reaches the discontinuity of the segment $R5$, increasing the periodicity by two, and the number of bounces as well. There seems to be no reason why this procedure will stop repeating itself, and so we will end up with a ``period increment bifurcation structure" \cite{gardini}. Points labeled $b$ and $c$ present the same feature, but with periodicities increasing in steps of three, and it seems there are many of such points in the bifurcation diagram. Three black points in figure \ref{fig8} indicate the experimental verifications of three rhythms that occurred in the regions defined by the period increment BBB labeled with $a$. These are $5:3$, $3:2$ and $1:1$. Again, these experimental findings where found with no difficulty and without passing through a transient behavior. 

\section{Discussion}

We have found the Biparametric Bifurcation Diagram with biphasic perturbation for the Tantalus oscillator, that has proven to be and ideal experimental model for reseting theory, as its oscillatory behavior is stable for long periods of time and shows no transient behavior when perturbed. 
In this section we present a comparison with the bifurcation diagram with monophasic perturbations of the Tantalus, published in a previous paper \cite{Arce}.

Resembling the monophasic Phase Transition Curves (PTCs), biphasic PTCs show discontinuities, up to the lever of accuracy of both the experimental and the theoretical techniques used. 
Another similarity is that bifurcations appear by a border collision mechanism. 
Nevertheless, several Big Bang Bifurcations (BBB) of period increment nature where found in the biphasic bifurcation diagram, in contrast with the monophasic case where no increment BBB where found. 

As mentioned in the Introduction section, only two works previously published have studied the effect of biphasic perturbations, to the best of our knowledge. 
Although bifurcation diagrams where not presented in these publications, the response to perturbations show some of the features observed in the Tantalus oscillator studied here. 
In \cite{Hortensia}, an experimental and theoretical study, a density oscillator is subject to biphasic perturbations of high intensity, and modeled as a Rayleigh oscillator. 
Biphasic perturbations to the FHN oscillator, consisting of Gaussian pulses of variable intensities, are studied in \cite{Croisier}. 
As observed in our work with the Tantalus, there are wide coupling intervals where perturbations have no effect, in both the density and the FHN oscillators (see figures 3(c) and 7(b) in \cite{Hortensia}, and figure 10 in \cite{Croisier}). 

Biphasic perturbations produce auto bistabilities of rhythm $1:1$ in the Tantalus oscillator, something that has not been found with monophasic perturbations of the Tantalus or other oscillators \cite{Croisier}. As with the density and FHN oscillators, these bistabilities are related to the regions where the PTC is close to the identity. 
Localized small deflections of the PTC appear when intensities grow, called ``bumps" by Guevara \cite{Croisier}, and since periodic perturbations are equivalent to a vertical shift of the PTC, the identity is crossed by the PTC in more than one point, creating many fixed points in the mapping (see figure \ref{fig5}(A), figure 5(a) in \cite{Hortensia}, and figure 10 in \cite{Croisier}). 
This bistability appears only in a very narrow interval of the perturbation periods, as explained in \cite{Croisier}.

Bistable regions of rhythms $1:1$ and $n:n-1$, are another feature that distinguishes the bifurcation diagram of biphasic perturbations, which is not present in the monophasic case. 
A region of this type crosses the bifurcation diagram diagonally from a point of $T_p= 0.1$ and $I=0.9$ to a point of $T_p= 0.9$ and $I=0.1$.
Two points  within this region, indicated as open circles in figure \ref{fig8}, appear to be BBB of period increment nature.

\begin{figure}[t]
\centering
\includegraphics[width=0.9\columnwidth]{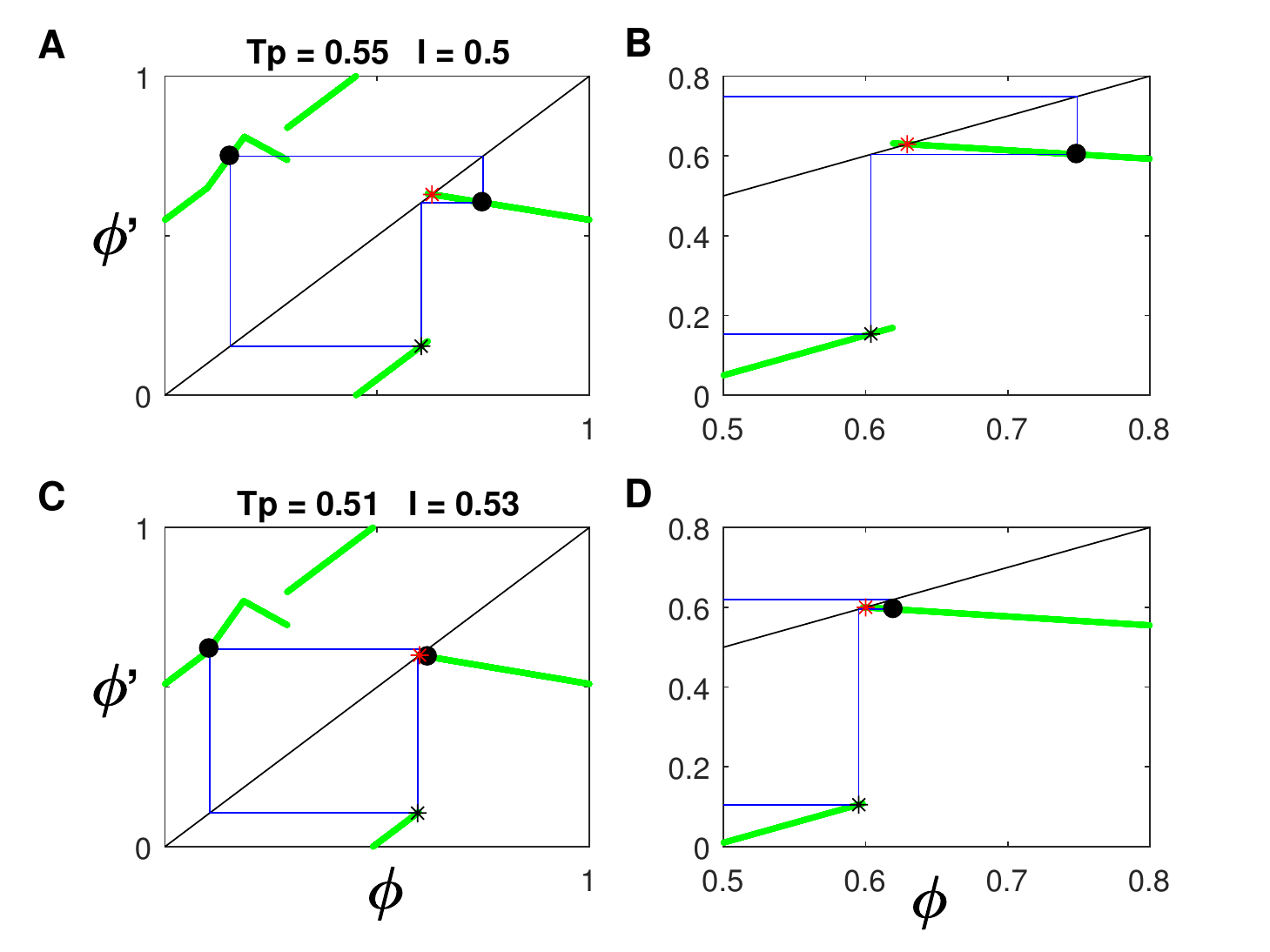}
\caption{Simultaneous border collision of the orbits of periodicities $1$ and $3$, at parameter values where a period increment BBB is found. In A, the red star indicates the point on the orbit of period $1$ that will collide from the right with the discontinuity, with a close up shown in B. The black star indicates the point where the orbit of period $3$ will collide with the discontinuity from the left. In C, an orbit with a selection of period perturbation and intensity such that both points (stars) are close to the discontinuity, with a close up shown in D.}
\label{fig10}
\end{figure}

Big Bang Bifurcations (BBB) have been studied extensively from a theoretical standpoint, specially by the group of Viktor Avrutin \cite{avrutin,gardini,avrutin2,avrutin3}. 
A particularly interesting feature of a BBB is that represent the point of convergence of orbits with an infinite number of different periodicities. 
Period increment BBB, like the ones found in the present work, are those in which the bifurcation parameter follow a sequence of the form $p = p_0 + nq$, for $q$ a given positive integer and $n= 1,2,3...$
A condition to have a BBB is found in the parameter space, and is provided by the crossing of two BCB bifurcation lines. 
This crossing implies that two stable orbits in the mapping bounce of a discontinuity, one coming from the left and the other from the right, regardless of the periodicities of the orbits.
For the BBB to be period increment, the slopes of the these orbits at the discontinuity must have opposite signs. 
Such orbits, one of period $1$ and another of period $3$ (like the one in figure \ref{fig9}(C) used to show a BCB), are shown in figure \ref{fig10} for the Tantalus oscillator under biphasic perturbations.
The fixed points that will collide with the discontinuity are indicated by a red and black asterisks for the the periods $1$ and $3$ respectively.
The remaining two point of the period $3$ orbit are black dots. 
Figure \ref{fig10}(A) shows the points of collision before reaching the discontinuity (see figure \ref{fig10}(B) for a closer look at the vicinity). 
Figure \ref{fig10}(C) shows the case where we got the colliding points the closest to the discontinuity (see figure \ref{fig10}(D) for a closer look), notice the slopes of opposite signs. 
The same behavior was found on the other two point indicated by open circles in figure \ref{fig8}. 
Period increment bifurcations have been found in other systems with an increment $q=1$. 
We have found experimentally a sequence with $q=2$ and theoretically two sequences with $q=3$.

The question of the generality of the features found in the bifurcation diagram for the Tantalus arises. 
We believe the density oscillator is a suitable candidate to search for such similarities. 
Preliminary studies using perturbations of moderate intensities suggest that density oscillator PTCs have discontinuities.
The size of these discontinuities depends on the intensity of the biphasic perturbations, as seen in the Tantalus oscillator. 
Therefore, we conjecture the structure of the bifurcation diagram of the density oscillator will resemble that of the Tantalus, with the existence of period increment BBBs. 
If the latter were true, experimental verification of coupling rhythms could be performed for a wider range.
 
\section{Conclusions}

The effects of biphasic perturbations on the Tantalus oscillator were studied in this work.
Comparison with monophasic perturbations, studied previously \cite{Arce}, show some similarities, but also remarkable differences. 
In both cases the Phase Transition Curves presented discontinuities, up to the degree of accuracy of our experiments.
Bifurcations between different periodic behaviors are the result of boundary collisions under biphasic perturbations, another similarity with the monophasic case. 
In contrast, biphasic perturbations have no effect over a wide range of the coupling interval which produces autobistabilities between different rhythms $1:1$, something unseen in the monophasic case. 
We observed many bistabilities between rhythms $1:1$ and rhythms of the type $n:n-1$, that are not observed under monophasic perturbations. 
Another unique feature of the biphasic case is the appearance of several period increment Big Bang Bifurcations that organize general bifurcation diagram.   
We found increments in steps of two and three. Experiments confirm the case with increments in steps of two. The latter result represents one of the few examples where increments of steps different from one have been found experimentally.

\section*{Acknowledgments}

We acknowledge the support of Cesar Arzate in the design and construction of the water injection devise.
We are also thankful to Eduardo Sacrist\'an for his advise in relation to the arduino platform, to Cesar Zepeda and Jorge Pineda for his support with the computer system. Finally, we are grateful to Fernanda Chac\'on for the design of figure \ref{fig1}.

\bibliographystyle{plain}

\end{document}